\documentclass[5p,twocolumn,preprint]{elsarticle}
\biboptions{sort&compress}
\usepackage{graphicx}
\usepackage{dcolumn}
\usepackage{bm}
\usepackage{color}
\usepackage{amsmath}
\usepackage{amssymb}

\newcommand{\be}{\begin{equation}}
\newcommand{\ee}{\end{equation}}
\newcommand{\ba}{\begin{eqnarray}}
\newcommand{\ea}{\end{eqnarray}}
\newcommand{\bd}{\begin{displaymath}}
\newcommand{\ed}{\end{displaymath}}
\renewcommand{\vec}[1]{\mbox{\boldmath$#1$}}

\begin{document}
\title{Interferometry for rotating sources}
	

\author[UoB]{S.~Velle \corref{cor1}}
\ead{Sindre.Velle@uib.no}
\author[UoB,FUoM]{S.~Mehrabi Pari }
\ead{Sharareh.Mehrabi.Pari@gmail.com}
\author[UoB]{L.P.~Csernai }
\ead{Laszlo.Csernai@uib.no}

\cortext[cor1]{Corresponding author}
\address[UoB]{ Department of Physics and Technology, University of Bergen,
    Allegaten 55, 5007 Bergen, Norway}
\address[FUoM]{Department of Physics, Ferdowsi University of Mashhad,
	91775-1436 Mashhad, Iran}

\date{\today}

\begin{abstract}
The two particle interferometry method to determine the size of 
the emitting source after a heavy ion collision is extended.
Following the extension of the method to spherical expansion
dynamics, here we extend the method to rotating systems.
It is shown that rotation of a cylindrically 
symmetric system leads to modifications, which can be perceived 
as spatial asymmetry by the "azimuthal HBT" method.

We study an exact rotating and expanding solution of the fluid dynamical
model of heavy ion reactions. We consider a source that is azimuthally
symmetric in space around the axis of rotation, and discuss the features
of the resulting two particle correlation function. This shows the
azimuthal asymmetry arising from the rotation. We show that this
asymmetry leads to results similar to those given by spatially
asymmetric sources.
\end{abstract}


\maketitle

\section{Introduction}

Two particle interferometry was adapted to heavy ion physics around 1980,
and turned out to be a sensitive tool of determining the space-time
extent of the source of the emission in heavy ion collisions.
It was a significant early discovery that pion interferometry
of exploding sources modifies the determination of the radius
of the source \cite{SP}.
This made the analysis a valuable tool for studying the 
dynamic expansion of nuclear sources.

Heavy ion collisions with finite impact parameters create systems where 
we have a large net angular momentum in the initial state, which leads 
to a rotating \cite{CMS} and expanding fireball.
If the formed Quark-Gluon Plasma (QGP) has low viscosity  \cite{CKM},
one can expect 
new phenomena like rotation or turbulence, which shows up in form of 
a starting Kelvin-Helmholtz instability (KHI) \cite{CsSA}. 
Rotation in heavy ion collision recently has been considered, and here 
we study 
a new class of exact hydrodynamic solutions for three dimensional, rotating 
and expanding cylindrically symmetric fireball \cite{CsoN14,ReHM,BW}.

The created system in relativistic heavy-ion collisions is microscopic 
and short-lived so only the momentum spectrum of the emitted particles 
can be measured directly. However, the space-time 
structure of the collision region can be studied using 
Hanbury-Brown-Twiss interferometry \cite{DCF}. This technique uses 
two particle correlations \cite{SP}, to probe the space-time shape of the 
particle emission zone. 
The size and shape of reaction zone become thus accessible,
with the "azimuthal HBT" method \cite{VC,Wiedemann,LHW,HK,HLW,RL,GBL}.

The Differential Hanbury Brown and Twiss (DHBT) method has been introduced 
earlier in \cite{DCF}. Previously the method has been applied to high 
resolution Particle in Cell Relativistic (PICR) fluid dynamical model 
\cite{DCF2} results, and it was shown that rotation can be detected 
by this modified method. The same PICR model was also 
used to calculate the vorticity of the flow, and due to the 
equipartition between the spin and orbit rotation the polarization
of emitted particles was evaluated \cite{BCW2013}, and turned out to 
be significant.  This early prediction was verified recently
\cite{Lisa2015}, by the experimental study of $\Lambda$ and $\bar{\Lambda}$
polarization in Au+Au reactions in the energy range of 
$\sqrt{s_{NN}} = 7.7 - 39$ GeV/nucl. Furthermore the 
 $\Lambda$ and $\bar{\Lambda}$ polarizations pointed in the same direction
that verified the mechanical, equipartition origin of the polarization
in contrast to electromagnetic origin, which would have led to 
opposite polarizations for  $\Lambda$ and $\bar{\Lambda}$.

In this work we calculate two pion correlation function for a rotating 
and expanding QGP, formed in Pb+Pb collisions 
at $\sqrt{s_{NN}}$ = 2.76 TeV/nucl, 
and impact parameter b = 0.7 $b_{max}$, by using the exact hydro 
model \cite{CsoN14,ReHM} we determine the effect of rotation on the 
correlation function (CF) for detectors at different positions.
Finally we fit results by "azimuthal HBT" to extract the apparent size 
of the rotating system in different directions.

\section{Correlation Function}

We use a simple Exact Model \cite{CsoN14} for expanding and
rotating systems to demonstrate the sensitivity of the
two particle correlation method to diagnose rotation.  Both 
polarization 
\cite{XGCs2015} 
and the two particle correlation 
\cite{DCF3} 
were already evaluated for this model, with parametrizations
adapted for peripheral heavy ion collisions
\cite{ReHM}.

We consider an azimuthally symmetric system around the 
rotation axis, $y$,  with Gaussian density profiles with
characteristic radii, $R$ and $Y$ and constant temperature
$T$. The initial parameters are given in Table \ref{t1}, and the
initial temperature is taken to be $T = 200$ MeV. 

The source function, $S(x,k)$, giving the emission rate
in the phase space, $x,k$, should be integrated over all points,
$x$, of the emitting source to obtain the correlation function:

\be
\begin{split}
& \int d^4x S(x,k) \propto \int w_s \gamma_s 
\left(k_0 + \vec{k} \cdot \vec{v_s}\right) \times \\ 
& \exp\left[-\frac{\gamma_s}{T_s}(k_0-\vec{k} \cdot \vec{v_s}) \right] 
 e^{-s_\rho/2}e^{-s_y/2} 
\frac{ds_y ds_\rho d\varphi}{\sqrt{(s_y)}} \ .
\end{split}
\label{spd}
\ee

Here the spatial integral is performed in cylindrical coordinates,
$s_y, s_\rho, \varphi$, where $s_y$, and $s_\rho$ are scaling 
variables, $s_y = y^2/Y^2$ and $s_\rho = (x^2+z^2)/R^2$.

The correlation function was evaluated the same way as in Ref.
\cite{DCF3}.
We should see that the source function explicitly depends
on the velocity field,  $\vec{v_s}$, which includes both the
expansion and the rotation of the system.

As the rotation axis is the $y$-axis, the $[x,z]$ plane is the
reaction plane. Due to the azimuthal symmetry the radius of the 
system in the $[x,z]$
 plane is $R$.

According to the conventions of two particle correlation functions
in heavy ion physics, the $z$-axis is the beam axis and determines 
the 
LONG direction. In the transverse plane, the $x$-axis (the direction of 
impact parameter) is transverse to the beam direction. In this way 
the 
OUT direction is the  $x$-direction. The remaining $y$-axis determines 
the 
SIDE direction. Due to the azimuthal symmetry of our specific
model the results for the LONG and OUT directions should be 
identical. 

The velocity field in $x, y, z$ coordinates is given by
\be
\begin{split}
& \vec v_s = \left(
\dot{R} \sqrt{s_\rho}\ \sin(\varphi) + 
R\omega \sqrt{s_\rho}\ \cos(\varphi), \right. \\ 
& \left. \dot{Y} \sqrt{s_y},
\dot{R} \sqrt{s_\rho}\ \cos(\varphi) - 
R\omega \sqrt{s_\rho}\ \sin(\varphi) \right)\ ,
\end{split}
\label{vel}
\ee 

where $\omega$ is the angular velocity, and $\varphi$ is the angle 
of rotation.

The mean transverse radius is $R=\sqrt{XZ}$, and we use this 
value when the exact model is studied.

\begin{table}[ht]
\begin{tabular}{ccccccc} \hline\hline \phantom{\Large $^|_|$}
	$t$  & $Y$ & $\dot{Y}$ & $\omega$ & $R$ & $\dot{R}$ & $\varphi$ \\
	(fm/c) & (fm) & (c)   &  (c/fm) &  (fm)   &   (c)    & (Rad) \\
	\hline
	0.   & 4.000   & 0.300 & 0.150   &  2.500   &  0.250    & 0.000 \\
	3.   & 5.258   & 0.503 & 0.059   &  3.970   &  0.646    & 0.307 \\
	8.   & 8.049   & 0.591 & 0.016   &  7.629   &  0.779    & 0.467 \\
\hline
\end{tabular}
\caption{
	Time dependence of characteristic parameters of the
	fluid dynamical calculation presented in Ref. \cite{ReHM}.  $R$ is
	the average transverse radius, $Y$ is the longitudinal length of the
	participant system, $\varphi$ is the angle of the rotation
        of the interior region of the system,
        around the $y$-axis, measured from the horizontal,
	beam, $z$-direction in the reaction, $[x,z]$, plane,
	$\dot{R}$ and $\dot{Y}$ are the speeds of expansion in transverse and
	longitudinal directions, and $\omega$ is the angular velocity of the
	internal region of the matter.
}
\label{t1}
\end{table}

In the practical calculations 
we use a detectors placed at 
$\vec k^+ = (k_x, k_y, k_z)  k = (0.924,0,0.383) k$, 
$\vec k = (1,0,0) k$ 
and 
$\vec k^- = (0.924,0,-0.383) k$,
which are orthogonal to the rotation axis, $\vec y$.

\section{Results}

\begin{figure}[ht] 
	\begin{center}
		\includegraphics[width=6cm]{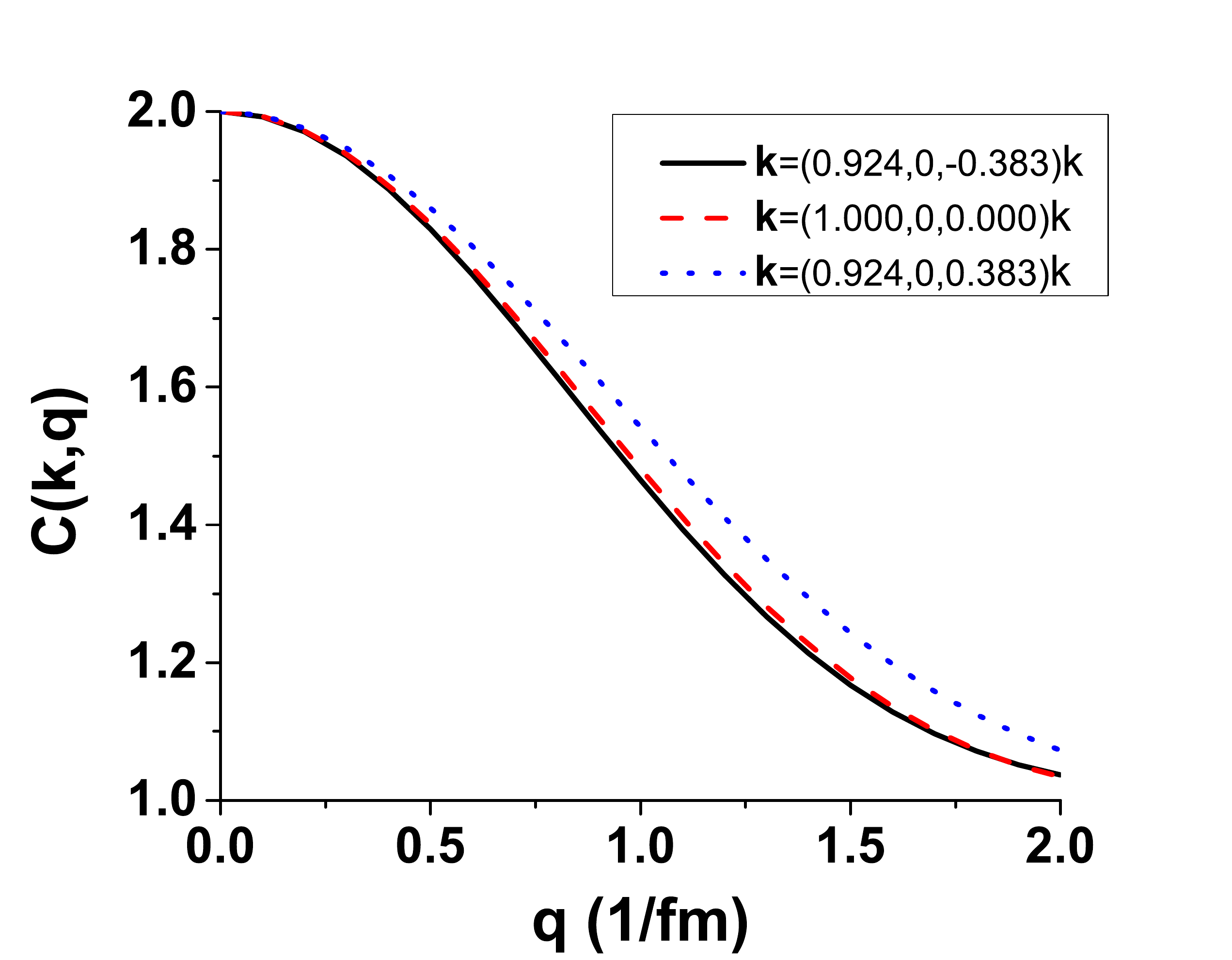}
	\end{center}
	\vskip -4mm
	\caption{ (color online)
		Correlation Function, $C(k,q)$, for the exact hydro model
                for the $q = q_{OUT}$ direction.
		R = 2.50 fm, $\dot{R}$ = 0.25 c, Y = 4.00 fm, 
		$\dot{Y}$ = 0.30 fm, $\omega$ = 0.30 c/fm, 
		at t = 0.0 fm/c with k = 5 fm$^{-1}$.	
		The solid black line is for measuring 
		the correlation function at $\vec k^- =(0.924,0,-0.383)k$, 
		the dashed red line is for $\vec k =(1,0,0)k$ 
		and the dotted blue line is for $\vec k^+ =(0.924,0,0.383)k$.}
	\label{ckq}
\end{figure}

\begin{figure}[ht] 
	\begin{center}
		\includegraphics[width=6cm]{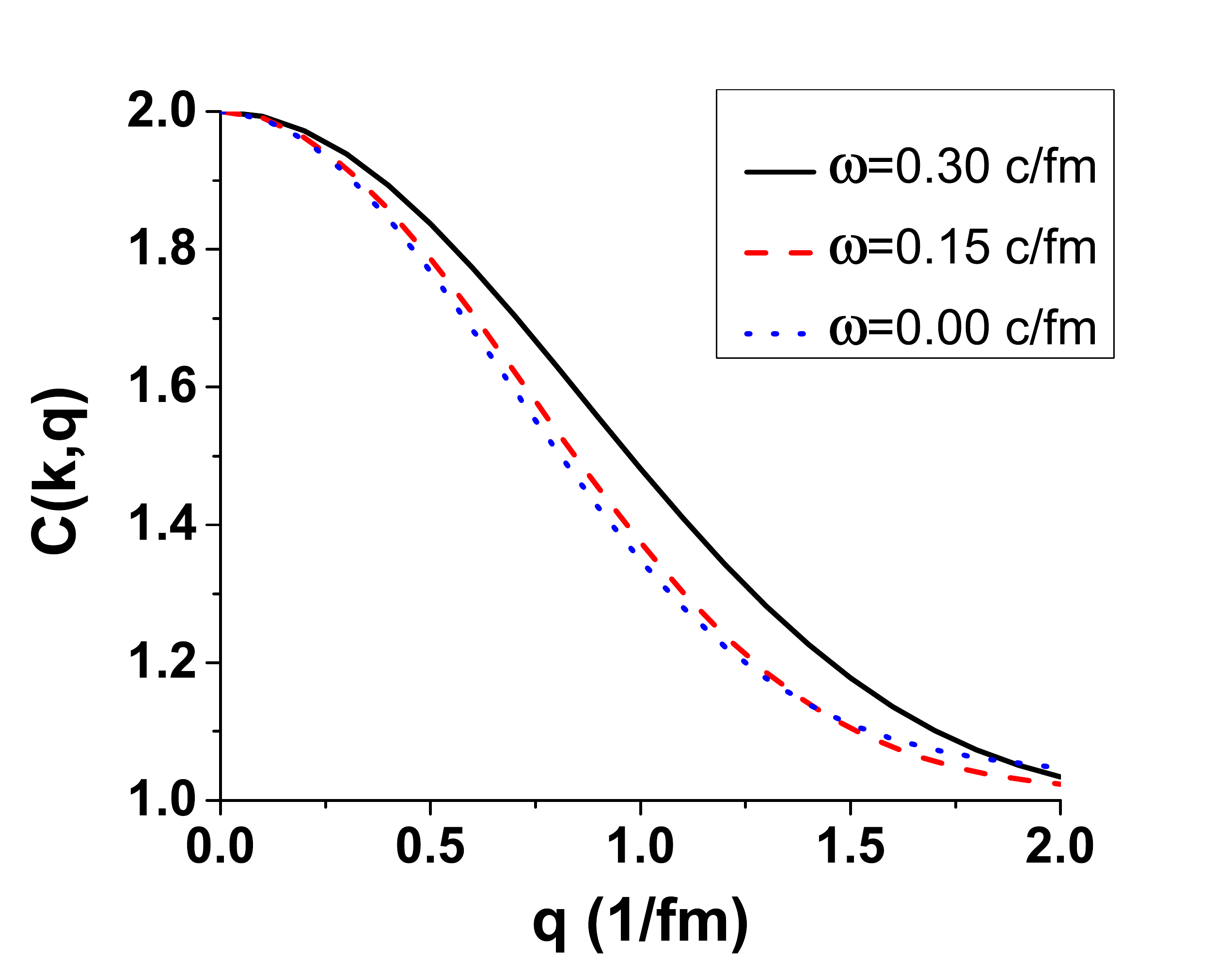}
	\end{center}
	\vskip -4mm
	\caption{ (color online)
		Correlation Function, $C(k,q)$, for the exact hydro model,
                for the $q = q_{OUT}$ direction.
		$R$ = 2.50 fm, $\dot{R}$ = 0.25 c, $Y$ = 4.00 fm, 
		$\dot{Y}$ = 0.30 fm at t = 0.0 fm/c with k =  5 fm$^{-1}$.	
		The solid black line is for $\omega=0.30$ c/fm, 
		the dashed red line is for $\omega=0.15$ c/fm 
		and the dotted blue line is for $\omega=0.00$ c/fm.}
	\label{ckq2}
\end{figure}

We calculate the CF for different values of 
the angular velocity, $\omega$, to see how it is affected. The CFs 
are shown in Figs. \ref{ckq} and \ref{ckq2}.

Subsequently the correlation functions is fitted by the azimuthal HBT 
method and parametrization:

\be
C(q,k)=1+exp\left(- \sum\limits_{i,j=L,O,S} q_i q_j R^2_{ij}(k) \right),
\label{ahbt}
\ee

The obtained radius parameters in the directions $L$ and $O$ are identical
due to the symmetry of the model, thus $R_{OO} = R_{LL}$.
We  took different $k$-values.

\begin{figure}[ht] 
	\begin{center}
		\includegraphics[width=9cm]{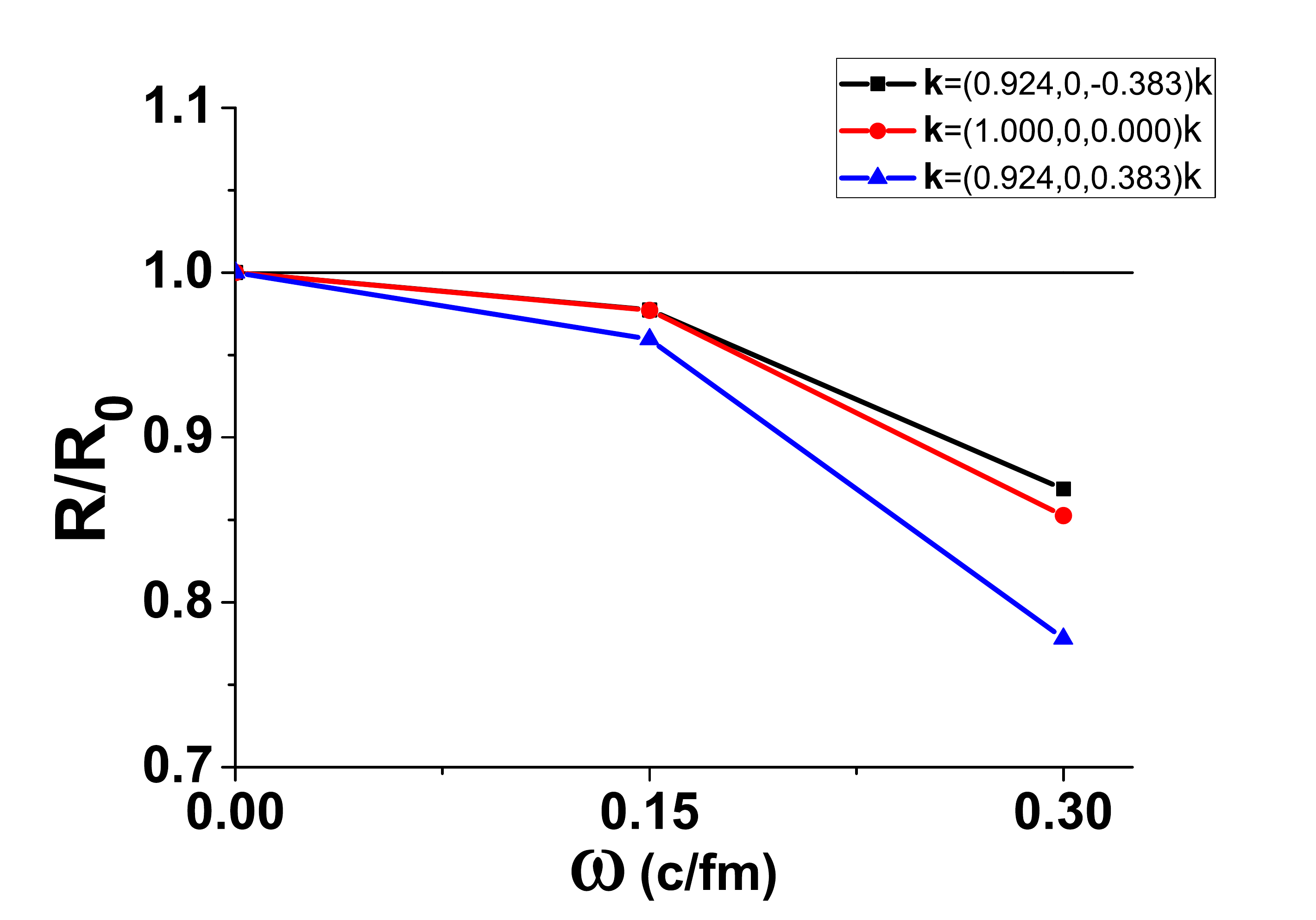}
	\end{center}
	\vskip -4mm
	\caption{ (color online)
		Ratio of radius from the fit for the correlation 
		function in Fig. \ref{ckq} and \ref{ckq2} 
        for different directions as a function of $\omega$, 
		the black line is for $\vec k^- = (0.924,0,-0.383)k$, 
		the red line is for $\vec k = (1,0,0)k$ and 
		the blue line is for $\vec k^+ = (0.924,0,0.383)k$.
		$R_0$ is the observed radius of the system without rotation.}
	\label{fit}
\end{figure}

We can compare the different radius parameters, $R^2_{ij}$, 
obtained by fitting the 
results of the CF obtained from the rotating and azimuthally
symmetric system, to the "azimuthal HBT" parametrization of eq. (\ref{ahbt}).
The obtained values,   $R \equiv R_{OO} = R_{LL}$, are shown
in Fig. \ref{fit}.

The CF increases for larger values of $\omega$ which corresponds to 
a decrease in the measured radius. The approximate radius decrease 
for $\omega=0$ to 0.15 c/fm and $\omega=0$ to 0.30 c/fm 
is 3-4\% and 15\% respectively.

In Fig. \ref{fit} we can see how the measured radius 
depends on the rotation velocity $\omega$ at time $t=0$ fm/c.. 
For a later times we would see a similar effect.
Even though $\omega$ becomes smaller for later times, 
the expansion velocity and size of the system are both influencing the CF
together with the rotation.
On the other hand there is no effect on the radius parameters
if either the rotation or expansion is zero \cite{DCF3}.

A higher rotation velocity will decrease the measured size of the system,
it also decreases more rapidly for larger values of $\omega$ as 
can be seen from the slope going
from $\omega=0$ to 0.15 c/fm and $\omega=0.15$ to 0.30 c/fm in Fig. \ref{fit}.
Asymmetry in the size is present if measured at 
different directions if the system
is rotating. If the rotation were reversed the 
correlation function will also change,
where the black and blue lines in Figs. \ref{ckq} and \ref{fit} 
are exchanged.

The detector at $k^+=(0.924,0,0.383)k$ shows a smaller measured radius 
for the exact hydro model while the radius 
showed at $k^-=(0.924,0,-0.383)k$ is larger.
This is also dependent on expansion velocity, 
temperature and size of the system.
The axial size, $Y$, is not affected by the rotation.

Thus the model results show that rotation significantly influences the
HBT evaluation similarly like the expansion, which influences data
significantly, e.g. in publications \cite{SWL09, KBS07, PW08, ZRW06, S05}.

\section{Conclusion}

Different values of the angular velocity will change the measured 
"azimuthal HBT"
size parameters of the system. It will also create smaller and larger values 
for the correlation function when measuring at different directions.
That the cylindrically symmetric system is rotating will be observed
as an asymmetric object in "azimuthal HBT" analysis.
\bigskip

\section*{Acknowledgements}

This work is supported by the Research Council of Norway, Grant no. 231469.


\end{document}